%% file: paper.tex
\documentclass{article}

\advance\textheight by 2cm
\advance\topmargin -1cm

\advance\textwidth by 2cm
\advance\oddsidemargin by -1cm
\advance\evensidemargin by -1cm

\usepackage[english]{babel} 
\usepackage[latin1]{inputenc}
\usepackage{pdfsync}
\usepackage{graphicx}
\usepackage{amsfonts,amsmath,mathrsfs,theorem}
\usepackage{rotating}
\usepackage{algorithm, algorithmic}
\floatname{algorithm}{Algorithm}
\usepackage[usenames,dvipsnames]{color}
\usepackage{multirow}
\usepackage{dcolumn}

\renewcommand{\epsilon}{\varepsilon}

\renewcommand{\theta}{\vartheta}

\def\..{\,\mathpunct{\ldotp\ldotp}} 

\newcommand{\comment}[1]{}

\begin{document}

\title{Robustness of Social Networks:\\{\LARGE Comparative Results
Based on Distance Distributions}}
 \author{Paolo Boldi \qquad Marco Rosa \qquad Sebastiano Vigna\\\normalsize
 Dipartimento di Scienze dell'Informazione\\\normalsize Universit\`a degli Studi
 di Milano, Italia}
\date{}
\bibliographystyle{alpha}

\maketitle

\begin{abstract}
Given a social network, which of its nodes have a stronger impact in determining
its structure? More formally: which node-removal order has the greatest impact on
the network structure? We approach this well-known problem for the first time in
a setting that combines both web graphs and social networks, using datasets that are orders of magnitude larger than those appearing in the previous literature, 
thanks to some recently
developed algorithms and software tools that make
it possible to approximate accurately the number of reachable pairs and the
distribution of distances in a graph. Our experiments highlight deep differences
in the structure of social networks and web graphs, show significant
limitations of previous experimental results, and at the same time reveal 
\emph{clustering by label propagation} as a new and very effective way of
locating nodes that are important from a structural viewpoint.
\end{abstract}

\section{Introduction}

In the last years, there has been an ever-increasing research activity in the
study of real-world complex networks~\cite{WFSNAMA} (the
world-wide web, the Internet autonomous-systems graph, coauthorship graphs,
phone call graphs, email graphs and biological networks, to cite a few). 
These networks, typically generated directly or indirectly by human activity and
interaction, appear in a large variety of contexts and often exhibit a surprisingly similar 
structure. One of the most important notions that researchers have been trying to capture
is ``node centrality'':
ideally, every node (often representing an
individual) has some degree of influence or importance within the social domain
under consideration, and one expects such importance to be reflected in the
structure of the social network; centrality is a quantitative measure that
aims at revealing the importance of a node.

Among the types of centrality that have been considered in the
literature (see \cite{BorCNF} for a good survey), many have to do with shortest
paths between nodes; for example, the \emph{betweenness centrality} of a node
$v$ is the sum, over all pairs of nodes $x$ and $y$, of the fraction of shortest
paths from $x$ to $y$ passing through $v$. The role played by shortest paths is
justified by one of the most well known features of complex networks, the so-called small-world phenomenon.    

A small-world network~\cite{CHCNSRF} is a graph where the average distance
between nodes is logarithmic in the size of the network, whereas
the clustering coefficient is large (that is, neighbourhoods tend to be denser)
than in a random Erd\H os-R\'enyi graph with the same size and average distance.\footnote{The
reader might find this definition a bit vague, and some variants are often
spotted in the literature: this is a general problem, also highlighted recently
in~\cite{LADTTSFG}.} Here, and in the following, by ``distance'' we mean the length of the shortest path between two nodes. The fact that social networks (either electronically mediated or not) exhibit the small-world property is known at least since Milgram's famous experiment~\cite{MilSWP}
and is arguably the most
popular of all features of complex networks.

Based on the above observation that the small-world property is by far
the most crucial of all the features that social networks exhibit, it is quite
natural to consider centrality measures that are based on node distance, like
betweenness. On the other hand, albeit interesting and profound, such measures
are often computationally too expensive to be actually computed on real-world
graphs; for example, the best known algorithm to compute betweenness
centrality~\cite{BraFABC} takes time $O(nm)$ and requires space for $O(n+m)$
integers (where $n$ is the number of nodes and $m$ is the number of arcs): both
bounds are infeasible for large networks, where typically $n \approx 10^9$ and $m
\approx 10^{11}$. 
For this reason, in most cases other strictly local measures of centrality are
usually preferred (e.g., degree centrality).  

One of the ideas that have emerged in the literature is that node centrality can
be evaluated based on how much the removal of the node ``disrupts'' the graph
structure~\cite{AJBEATCN}. This idea provides also a notion of robustness
of the network: if removing few nodes has no noticeable impact, then the network
structure is clearly robust in a very strong sense. On the other hand, a
node-removal strategy that quickly affects the distribution of distances
probably reflects an importance order of the nodes.

Previous literature has used mainly the diameter or some analogous measure to
establish whether the network structure changed. Recently, though, there have
been some successful attempts to produce reliable estimates of the
\emph{neighbourhood function} of very large graphs~\cite{PGFANF,BRVH}; an
immediate application of these approximate algorithms is the computation of the
number of \emph{reachable pairs} of the graph (the number of pairs $\langle x,
y\rangle$ such there is a directed path from $x$ to $y$) and its \emph{distance
distribution} (the distance distribution of a graph is a discrete distribution
that gives, for every $t$, the fraction of pairs of nodes that are at distance
$t$). From this data, a number of existing measures can be computed quickly and
accurately, and new one can be conceived.

We thus consider a certain ordering of the nodes of a graph (that is supposed to
represent their ``importance'' or ``centrality''). We remove nodes (and of
course their incident arcs) following this order, until a certain percentage
$\theta$ of the arcs have been deleted\footnote{Observe that we delete nodes but
count the percentage of arcs removed, and not of nodes: this choice is justified
by the fact that otherwise node orderings that put large-degree nodes first
would certainly be considered (unfairly) more disruptive.}; finally, we compare
the number of reachable pairs and distance distribution of the new graph with
the original one. The chosen ordering is considered to be a reliable
measure of centrality if the measured difference increases rapidly with $\theta$
(i.e., it is sufficient to delete a small fraction of important nodes to change
the structure of the graph).

In this work, we applied the described approach to a number of complex networks,
considering different orderings, and obtained the following results:
\begin{itemize}
 \item In all complex networks we considered, the removal of a limited fraction
 of randomly chosen nodes does not change the distance distribution
 significantly, confirming previous results.
 \item We test strategies based on PageRank and on clustering (see
Section~\ref{sec:strategies} for more information about this), and show that
they (in particular, the latter) disrupt quickly the structure of a web graph.
 \item Maybe surprisingly, none of the above strategies seem to have an impact
 when applied to social networks other than web graphs. This is yet another example
 of a profound structural difference between web graphs and social
 networks,\footnote{We remark that several proposals have been made to find
 features that highlight such structural differences in a
 computationwise-feasible way (e.g., assortative mixing~\cite{NePWSNDOTN}),
 but all instances we are aware of have been questioned by the subsequent
 literature, so no clear-cut results are known as yet.} on the
 same line as those discussed in~\cite{BRVH} and~\cite{CKLCSN}. This observation, 
 in particular, suggests that social networks tend to be much more robust and 
 cohesive than web graphs, at least as far as distances are concerned, and that
 ``scale-free'' models, which are currently proposed for both type of networks,
 do not to capture this important difference.
\end{itemize}

\section{Related work}

The idea of grasping information about the structure of a network by
repeatedly removing nodes out of it is not new: Albert, Jeong and
Barab{\'a}si~\cite{AJBEATCN} study experimentally the variation of the 
diameter on two different models of \emph{undirected} random graphs when nodes
are removed either randomly or in ``connectedness order''
and
report different behaviours. They also perform tests on some small real data
set, and we will compare their results with ours in Section~\ref{sec:discussion}.

More recently, 
node-centrality measures that look at how some graph invariant changes when some 
vertices or edges are deleted (sometimes called ``vitality''~\cite{BENAMF} or
``induced'' measures) have been studied for example in~\cite{BorISKPSN}
(identifying nodes that maximally disconnect the network) or in~\cite{BCKRCMCID} (related to the uncertainty of
data).

Donato, Leonard, Millozzi and Tsaparas~\cite{DLMTMIWG} study how the size of the
giant component changes when nodes of high indegree or outdegree are removed from
the graph. While this is an interesting measure, it does not provide information
about what happens outside the component. They develop a library for
semi-external visits that make it possible to compute in an exact way the strongly connected
components on large graphs.

Finally, Fogaras~\cite{FogWSBW} considers how the
\emph{harmonic diameter}\footnote{Actually, the notion had been introduced
before by Marchiori and Latora and named \emph{connectivity
length}~\cite{MaLHSW}, but we find the name ``harmonic diameter'' much more
insightful.} (the harmonic mean of the distances) changes as nodes are deleted from a small (less than one million node) snapshot of the \texttt{.ie} domain, reporting a large increase (100\%) when as little as 1000 nodes with high PageRank are removed. The harmonic diameter is estimated by a small number
of visits, however, which gives no statistical guarantee on the accuracy of the
results.

Our study is very different. First of all, we use graphs that are two
orders of magnitude larger than those considered in~\cite{AJBEATCN} or~\cite{FogWSBW}; moreover,
we study the impact of node removal on the whole spectrum of distances. Second,
we apply removal procedures to large social networks
(previous literature used only web or Internet graphs), and the striking difference in
behaviour shows that ``scale-free'' models fail to capture essential differences between these kind of networks and web graphs.
Third, we document in a reproducible way all our experiments, which have provable statistical
accuracy.

\section{Computing the distance distribution}

Given a directed graph $G$, its \emph{neighbourhood function} $N_G(t)$ 
returns for each $t\in\mathbf N$ the number of pairs of nodes $\langle x, y\rangle$ such that
$y$ is reachable from $x$ in no more than $t$ steps. From the neighbourhood
function, several interesting features of a graph can be estimated, and in this
paper we are especially interested in the \emph{distance distribution} of the
graph $G$ , represented by the cumulative distribution function
$H_G(t)$, which returns the fraction of reachable pairs at distance at most
$t$, that is, $H_G(t) = {N_G(t)}/{\max_t N_G(t)}$.
The corresponding probability density function will be denoted by $h_G(-)$.



Recently, HyperANF~\cite{BRVH} emerged as an evolution of the ANF
tool~\cite{PGFANF}. HyperANF can compute for the first time in a few hours the
neighbourhood function of graphs with billions of nodes with a small error and
good confidence using a standard workstation.
%
%
The free availability of
HyperANF opens new and interesting ways to study large graphs, of which this
paper is an example.



\section{Removal strategies and their analysis}

In the previous section, we discussed how we can effectively approximate the
distance distribution of a given graph $G$; we shall use such a
distribution as the graph structural property of interest. 

Consider now a given total order $\prec$ on the nodes of $G$; we think of
$\prec$ as a removal strategy in the following sense: when we want to remove
$\theta m$ arcs, we start removing the $\prec$-largest node (and its
incident arcs), go on removing the second-$\prec$-largest node etc.
and stop as soon as $\geq\theta m$ arcs have been removed. The resulting graph
will be denoted by $G(\prec,\theta)$. Of course, $G(\prec,0)=G$ whereas
$G(\prec,1)$ is the empty graph. We are interested in applying some measure of
\emph{divergence}\footnote{We purposedly use the word ``divergence'' between
distributions, instead of ``distance'', to avoid confusion with the notion of
distance in a graph.} between the distribution $H_G$ and the distribution
$H_{G(\prec,\theta)}$. By looking at the divergence when $\theta$
varies, we can judge the ability of $\prec$ to identify nodes that will disrupt
the network.

\subsection{Some removal strategies}
\label{sec:strategies}

We considered several different strategies for removing nodes from a graph. Some
of them embody actually significant knowledge about the structure of the graph, whereas
others are very simple (or even independent of the graph) and will be used as
baseline. Some of them have been used in the previous literature, and will be
useful to compare our results.

As a first observation, some strategies requires a symmetric graph
(a.k.a.,~undirected). In this case, we symmetrise the graph by adding the missing arcs\footnote{It is
mostly a matter of taste whether to use directed symmetric graphs or simple
undirected graphs. In our case, since we have to cope with both directed and
undirected graph, we prefer to speak of directed graphs that are symmetric,
that is, for every arc $x\to y$ there is a symmetric arc $y\to x$.}.

The second obvious observation is that some strategies might depend on available
metadata (e.g., URLs for web graphs) and might not make sense for all graphs.

\begin{description}
\item[Random.] No strategy: we pick random nodes and remove them from the graph.
It is important to test against this ``nonstrategy'' as we can show that the
phenomena we observe are due to the peculiar choice of nodes involved, and not
to some generic property of the graph.
\item[Largest-degree first.] We remove nodes in decreasing (out)degree order.
This strategy is an obvious baseline, as \emph{degree centrality} is the
first shot at centrality in a network.
\item[Near-Root.] In web graphs, we can assume that nodes that are roots of web
sites and their (quasi-)immediate successors (e.g., pages linked by the
root) are most important in establishing the distance distribution, as people
tend to link higher levels of web sites. This strategy removes essentially first root nodes, then the nodes
that are children of a root on, and so on.
\item[PageRank.] PageRank~\cite{PBMPCR} is an well-known algorithm that assigns
ranks to nodes using a Markov chain based on the structure of the graph. It has
been designed as an improvement over degree centrality, because nodes with high
degree which however are connected to nodes of low rank will have a rather low
rank, too (the definition is indeed recursive). There is a vast body of
literature on the subject: see~\cite{BSVPFD,LaMeDIPR} and the references
therein.
\item[Label propagation.] Label propagation~\cite{citeulike:1724653} is a
powerful technique for clustering symmetric graphs. Each node has a label (initially, the node number itself) and
through a number of rounds each node changes its label by taking the label of the majority of its
neighbours. At the end, node labels are used as cluster identifiers. Our removal
strategy picks first, for each cluster in decreasing size order, the node with
the highest number of neighbours in other clusters: intuitively, it is a representative of a set of
tightly connected nodes (the cluster) which however has a very significant
connection with the outside world (the other clusters) and thus we expect that
its removal should seriously disrupt the distance distribution. Once we have
removed all such nodes, we proceed again, cluster by cluster, using the same
criterion (thus picking the second node of each cluster that has
more connection towards other clusters), and so on.
\end{description}

\subsection{Measures of divergence} 
\label{sec:measures}

Once we changed the structure of a graph by deleting some of its nodes (and
arcs), there are several ways to measure whether the structure of the graph has
significantly changed. The first, basic raw datum we consider is the
\emph{number of pairs of nodes that are still reachable} divided by \emph{the
number of pairs initially reachable}, expressed as a percentage. Then, to
estimate the change of the distance distribution we considered the following possibilities (here $P$
denotes the original distance distribution, and $Q$ the distribution after node
removal):
\begin{description}
\item[Relative average-distance change.] This is somehow the simplest and most
natural measure: how much has the average distance changed? We use the measure
\[
	\delta(P,Q)=\frac{\mu_Q}{\mu_P}-1
\]
where $\mu$ denotes the average; in other words, we measure how much the
average value changed. This
measure is non-symmetric, but it is of course easy to obtain $\delta(P,Q)$ from $\delta(Q,P)$.
\item[Relative harmonic-diameter change.] This measure is analogous to the
relative average-distance change, but the average on distances is \emph{harmonic} and
\emph{computed on all pairs}, that is:
\[
\frac{n(n-1)}{\sum_{x\neq y}\frac1{d(x,y)}} = n(n-1)\big/\sum_{t>0}
\frac1t(N_G(t)-N_G(t-1)),
\]
where $n$ is the number of nodes of the graph. This measure, used
in~\cite{FogWSBW}, combines reachability information, as unreachable pairs
contribute zero to the sum. It is easily computable from the neighbourhood
function, as shown above.
\item[Kullback-Leibler divergence.] This is a measure of \emph{information
gain}, in the sense that it gives the number of additional bits that are necessary to
code samples drawn from $P$ when using an optimal code for $Q$. Also this
measure is non-symmetric, but there is no way obtain the divergence from $P$ to
$Q$ given that from $Q$ to $P$.
\item[$\ell$ norms.] A further alternative is given by viewing distance
distributions as functions $\mathbf N\to[0\..1]$ and measure their distance
using some $\ell$-norm, most notably $\ell_1$ or $\ell_2$. Such distances are of
course symmetric. 
\end{description}

\begin{figure}
\centering
\begin{tabular}{c}
\includegraphics[scale=.6]{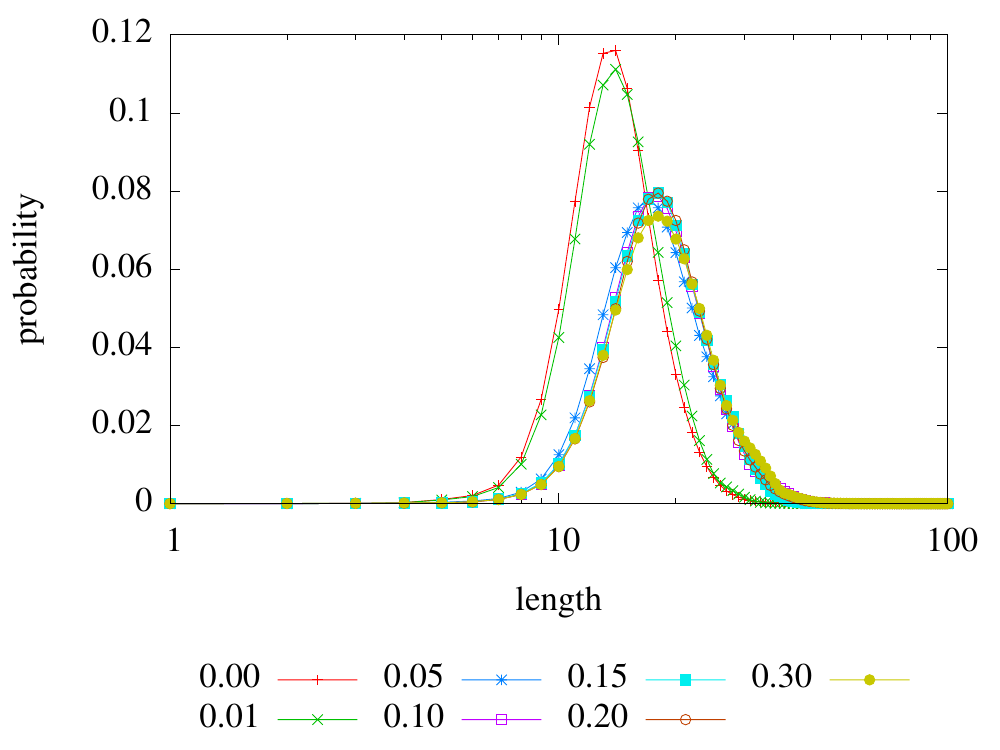}\includegraphics[scale=.6]{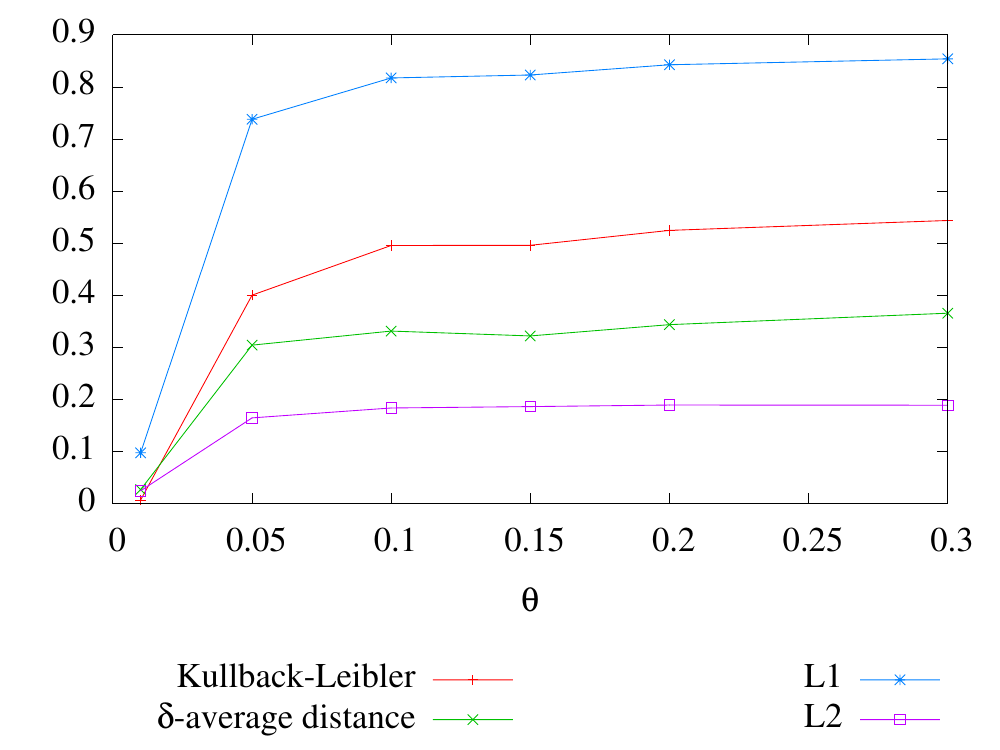}
\end{tabular}
\caption{\label{fig:divergences} Testing various divergence measures on a
web graph (a snapshot of the \texttt{.it} domain of 2004) and the near-root
removal strategy. You can see how the distance distribution
changes for different values of $\theta$ and the behaviour of divergence
measures. We omitted to show the harmonic-diameter change to make the plot
easier to read.}
\end{figure}

We tested, with various graphs and removal strategies, how the choice of
distribution divergence influences the interpretation of the results obtained.
In Figure~\ref{fig:divergences} we show this for a single web graph and a single
strategy, but the outcomes agree on all the graphs and strategies tested: the
interpretation is that all divergences agree, and for this reason we shall use
the (simple) measure $\delta$ applied to the average distance in the
experimental section. The advantage of $\delta$ over the other measures is that it is very easy to interpret; for example, if
$\delta$ has value, say, $0.3$ it means that node removal has increased the
average distance by $30\%$. We also discuss $\delta$ applied to the harmonic
diameter.

\section{Experiments}

For our experiments, we considered a number of networks with various sizes and
characteristics; most of them are either web graphs or (directed or undirected)
social graphs of some kind (note that for
web graphs we can rely on the URLs as external source of information). More
precisely, we used the following datasets:

\begin{itemize}
\item \emph{Hollywood}: 
One of the most popular \emph{undirected} social graphs, the graph of
movie actors: vertices are actors, and two actors are joined
by an edge whenever they appeared in a movie together. 
\item \emph{LiveJournal}:
LiveJournal
is a
virtual community social site started in 1999: nodes are
users and there is an arc from $x$ to $y$ if $x$ registered $y$ among his
friends (it is not necessary to ask $y$ permission, so the graph is
\emph{directed}). We considered the same 2008 snapshot of \emph{LiveJournal}
used in~\cite{CKLCSN} for their experiments
\item \emph{Amazon}: This dataset describes similarity among books as reported
by the Amazon store; more precisely the data was
obtained
in 2008 using the Amazon E-Commerce Service APIs using
\texttt{SimilarityLookup} queries.
\item \emph{Enron}:
This dataset was made public by the Federal Energy Regulatory Commission during
its investigations: it is a partially anonymised corpus of e-mail messages
exchanged by some Enron employees (mostly part of the senior management). We
turned this dataset into a \emph{directed} graph, whose nodes represent people
and with an arc from $x$ to $y$ whenever $y$ was the recipient of (at least) a
message sent by $x$.
\item For comparison, we considered two web graphs of different size: a 2004
snapshot of the \texttt{.it} domain ($\approx40$ million nodes), and a snapshot
taken in May 2007 of the \texttt{.uk} domain ($\approx100$ million nodes).
\end{itemize}

We remark that all our graphs are available at the LAW
web site.\footnote{\texttt{http://law.dsi.unimi.it/}. In particular, the graphs
we used are the datasets named \texttt{hollywood-2009},
\texttt{ljournal-2008}, \texttt{amazon-2008}, \texttt{enron}, \texttt{it-2004} and
\texttt{uk-2007-05}.} HyperANF is available as free software at the WebGraph web site\footnote{\texttt{http://webgraph.dsi.unimi.it/}}, 
and the class \texttt{RemoveHubs} that has been used to perform the experiments
we describe is part of the LAW software.

We applied our removal strategies with different impact levels (e.g., percentage
of removed arcs), namely $0.01$, $0.05$, $0.1$, $0.15$, $0.2$ and $0.3$. For
each level we ran HyperANF at least seven times using 128 registers per counter:
the percentage of reachable pair displayed in our tables has been obtained by averaging the neighbourhood
functions obtained from the runs, with relative standard deviation smaller than
$3.5\%$ (e.g., the measure is within relative error $10.5\%$ with $95\%$
confidence). The starting number of reachable pairs is known with relative standard
deviation smaller than $0.1\%$. The remaining derived measurements (average distances and harmonic
diameters) have been computed separately on each run, and the resulting relative standard
deviation is less than $4\%$ for the average distance, and less than $20\%$ for
the harmonic diameter, except for about a dozen measurements, where it is less
than $8.5\%$ for the average distance, and less than $30\%$ for the harmonic
diameter.\footnote{Unfortunately, estimating with precision the harmonic
diameter is difficult due to the nonlinearity of its definition.} Our
tables and graphs slightly differs from those previously published~\cite{BRVRSN} because we had time to generate more runs, and thus increase the precision of our results: some variation is also observed because of the relatively small number of runs (unavoidable, due to the large number of graphs to be analyzed).

\begin{sidewaysfigure}
\centering
\noindent\includegraphics[scale=.7]{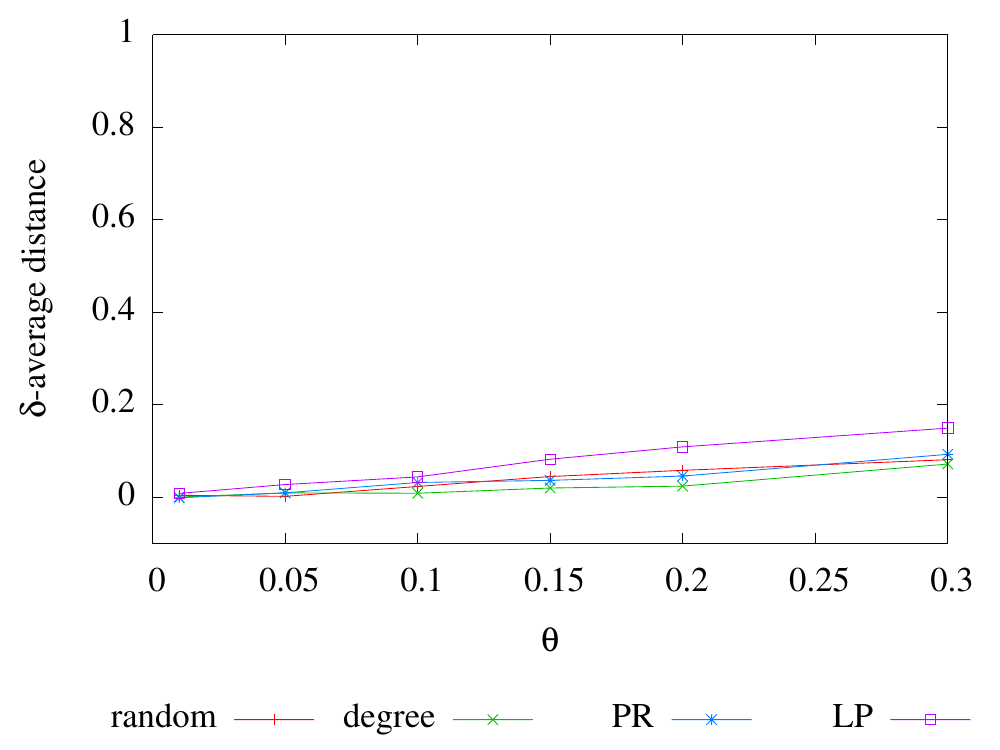}%
\includegraphics[scale=.7]{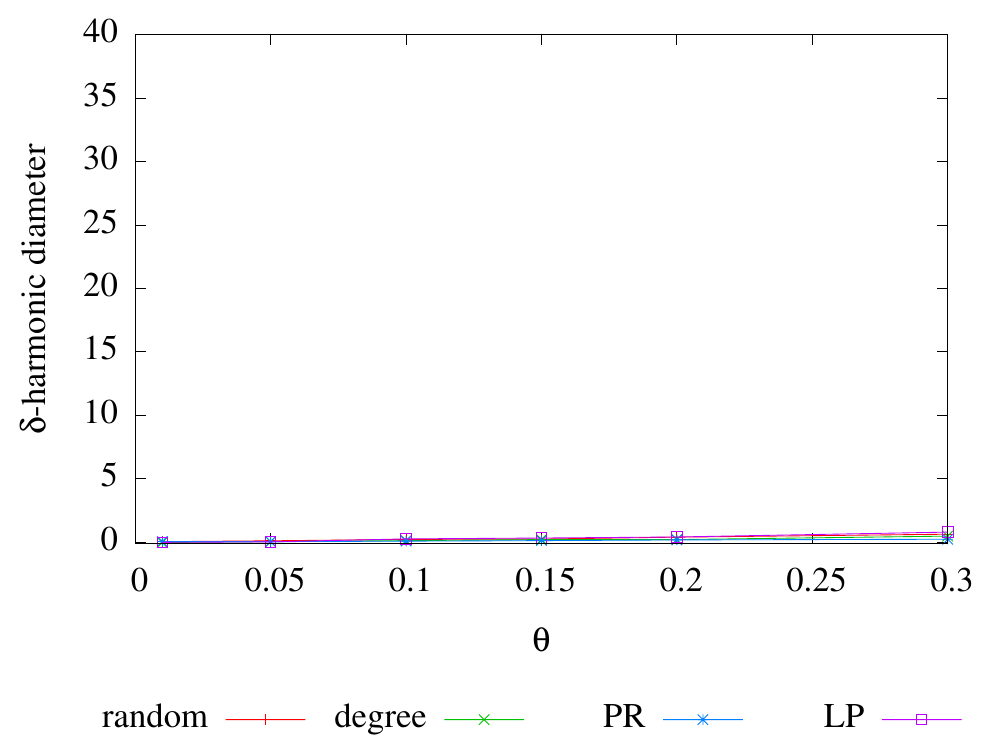}%
\includegraphics[scale=.7]{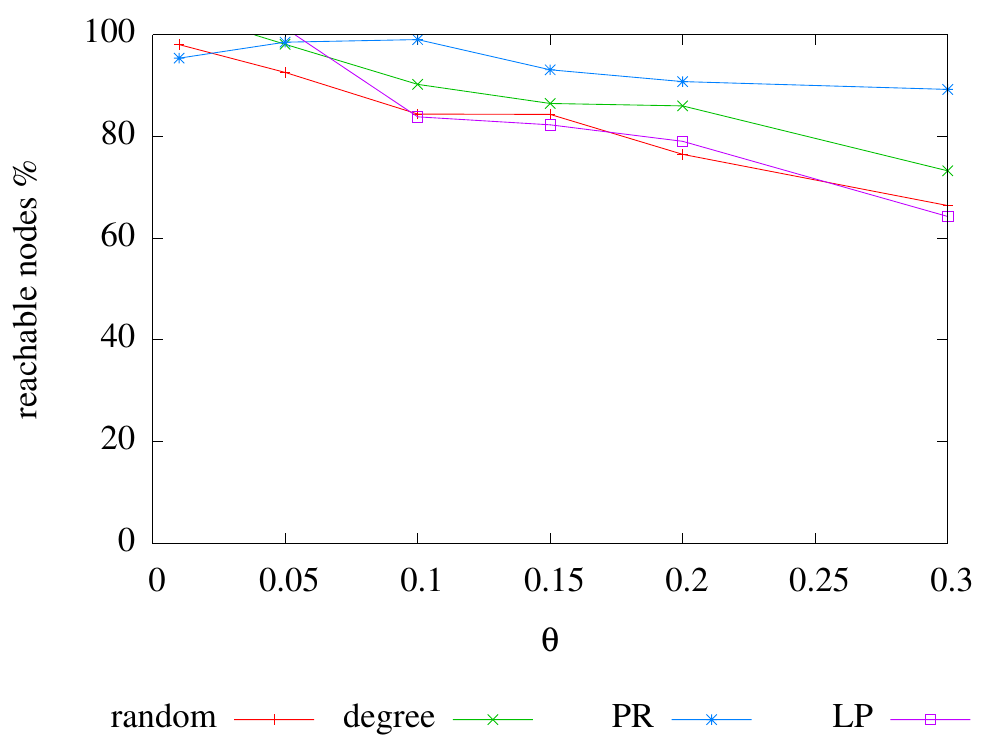}\\
Amazon\\[1em]
\noindent\includegraphics[scale=.7]{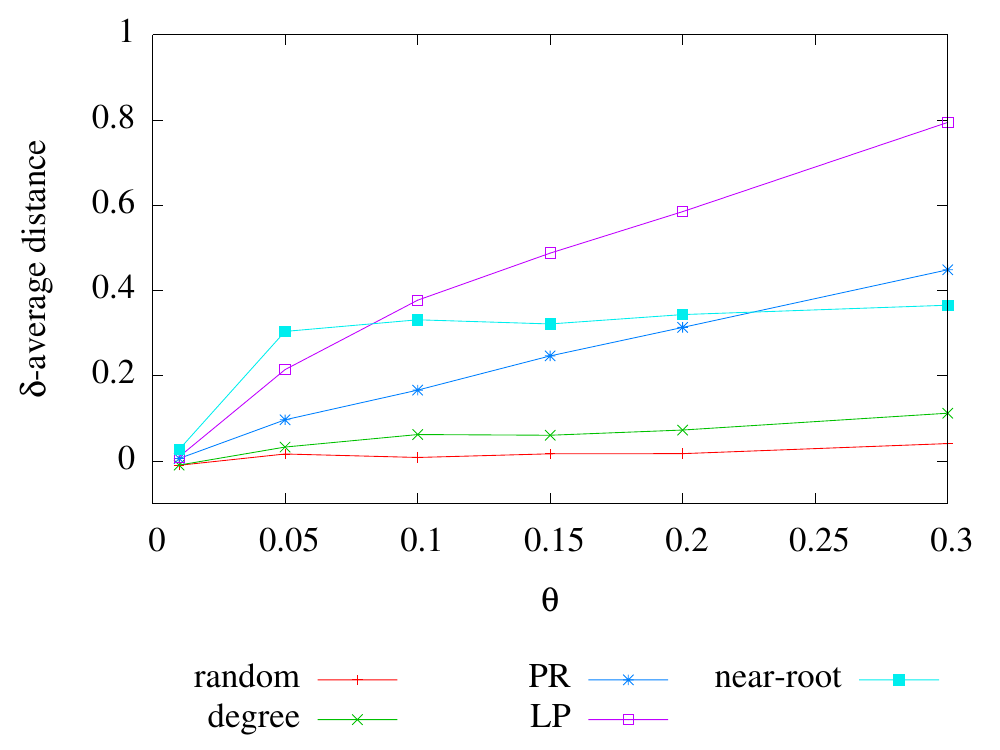}%
\includegraphics[scale=.7]{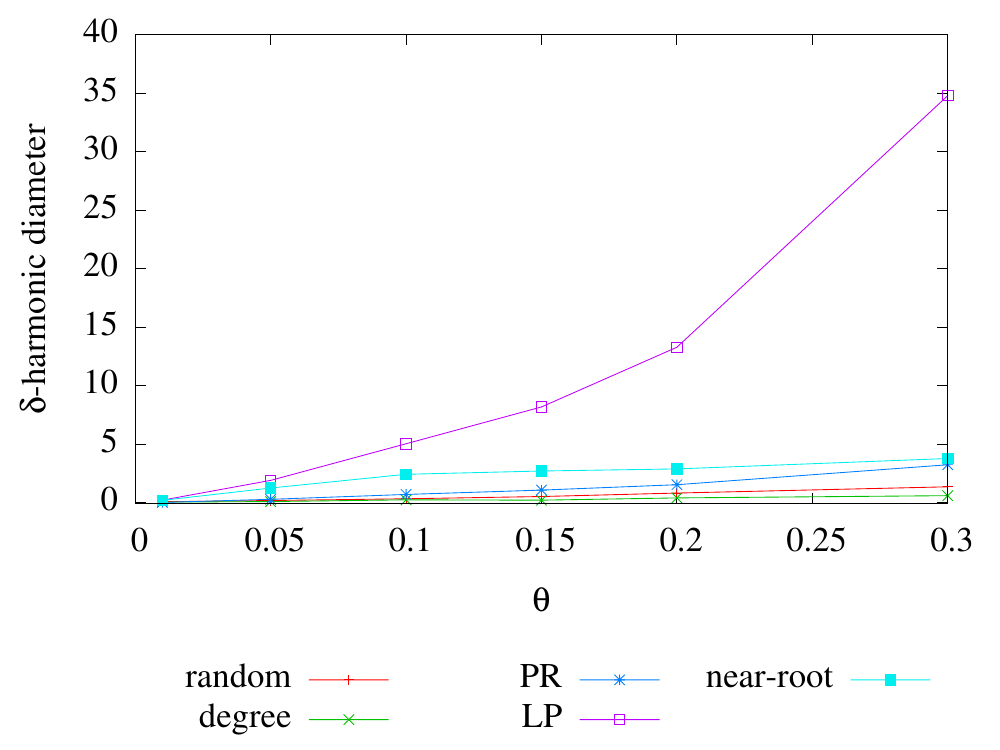}%
\includegraphics[scale=.7]{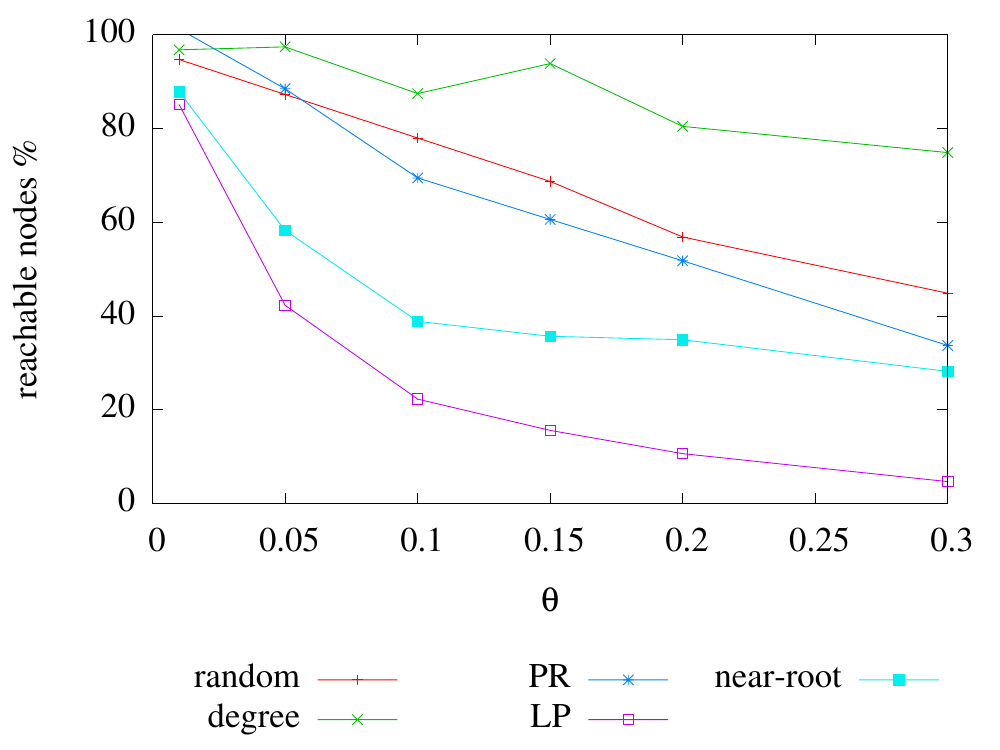}\\
\texttt{.it}
\caption{\label{fig:strategies} Typical behaviour of social networks
(Amazon, upper) and web graphs (\texttt{.it}, lower) when a $\theta$
fraction of arcs is removed using various strategies. None of the proposed
strategies completely disrupts the structure of social networks, but the
effect of the label-propagation removal strategy on web graphs is very visible.}
\end{sidewaysfigure}

\begin{sidewaystable}\centering
\begin{tabular}{|l|l|D{.}{.}{3}r|D{.}{.}{3}r|D{.}{.}{3}r|D{.}{.}{3}r|D{.}{.}{3}r|D{.}{.}{3}r|} \hline
	Graph 	& Strategy 	& \multicolumn{2}{|c|}{$0.01$} & \multicolumn{2}{|c|}{$0.05$} & \multicolumn{2}{|c|}{$ 0.1 $}  & \multicolumn{2}{|c|}{$ 0.15 $} & \multicolumn{2}{|c|}{$ 0.2 $} & \multicolumn{2}{|c|}{$ 0.3 $}   \\ \hline
\input avg.table
\end{tabular}
\vspace{.5em}
\caption{\label{tab:results}For each graph and a sample of fractions of removed
arcs we show the change in average distance (by the measure $\delta$ defined in
Section~\ref{sec:measures}) and the percentage of reachable pairs. PR stands
for PageRank, and LP for label propagation.}
\end{sidewaystable}

\section{Discussion}
\label{sec:discussion}

Table~\ref{tab:results} and Figure~\ref{fig:strategies} show that social
networks suffer spectacularly less disconnection than web graphs when their nodes are
removed using our strategies. Our most efficient removal strategy, label
propagation, can disconnect almost all pairs of a web graph by removing 30\% of
the arcs, whereas it disconnects only about half (or less) of the pairs on
social networks. This entirely different behaviour shows that web graphs have a path
structure that passes through fundamental hubs.


Moreover, the average distance of the web graphs we consider increases by
50$-$80\% upon removal of 30\% of the arcs, whereas in most social networks
there is just an increase of a few percents (in any case, always
less than $20\%$).\footnote{We remark that in some cases the measure is negative
or does not decrease monotonically. This is an artifact of the probabilistic technique used to estimate the number of
pairs---small relative errors are unavoidable.}

Note that random removal can
separate a good number of reachable pairs, but the increase in average distance
is very marginal. This shows that considering both measures is important in
evaluating removal strategies.

Of course, we cannot state that there is no strategy able
to disrupt social networks as much as a web graph (simply because this strategy
may be different from the ones that we considered), but the fact all strategies
work very similarly in both cases (e.g., label propagation is by far the most 
disruptive strategy) suggests that the phenomenon is intrinsic.

There is a candidate easy explanation: shortest paths in web graphs
pass frequently through home pages, which are linked more than other pages. But this
explanation does not take into account the fact that clustering by label
propagation is significantly more effective than the near-root removal
strategy. Rather, it appears that there are fundamental hubs (not necessarily
home pages) which act as shortcuts and through which a large number of shortest
paths pass. Label propagation is able to identify such hubs, and their
removal results in an almost disconnected graph and in a very significant
increase in average distance.

These hubs are not necessarily of high outdegree: quite the opposite, rather, is
true. The behaviour of web graphs under the largest-degree strategy is
illuminating: we obtain the smallest reduction in reachable pairs and an almost
unnoticeable change of the average distance, which means that nodes of high
outdegree are not actually relevant for the global structure of the network.

Social networks are much more resistant to node removal. There is no strict
clustering, nor definite hubs, that can be used to eliminate or elongate
shortest paths. This is not surprising, as networks emerging from social
interaction are much less engineered (there is no notion of ``site'' or ``page
hierarchy'', for example) than web graphs.

The second important observation is that the removal strategies based
on PageRank and label propagation are always the best (with the exception of
the near-root strategy for \texttt{.uk}, which is better than PageRank). This
suggests that label propagation is actually able to identify structurally
important nodes in the graph---in fact, significantly better than any 
other method we tested.

Is the ranking provided by label propagation correlated to other rankings?
Certainly not to the other rankings described in this paper, due to the
different level of disruption it produces on the network. The closest ranking
with similar behaviour is PageRank, but, for instance, Kendall's $\tau$ between
PageRank and ranking by label propagation on the \texttt{.uk} dataset is
$\approx -0.002$ (complete uncorrelation).

\begin{sidewaystable}\centering
\begin{tabular}{|l|l|D{.}{.}{3}r|D{.}{.}{3}r|D{.}{.}{3}r|D{.}{.}{3}r|D{.}{.}{3}r|D{.}{.}{3}r|} \hline
	Graph 	& Strategy 	& \multicolumn{2}{|c|}{$0.01$ } & \multicolumn{2}{|c|}{$0.05$ } & \multicolumn{2}{|c|}{$0.1$}  & \multicolumn{2}{|c|}{$0.15$} & \multicolumn{2}{|c|}{$0.2$} & \multicolumn{2}{|c|}{$0.3$} \\ \hline
\input harm.table
\end{tabular}
\vspace{.5em}
\caption{\label{tab:harmonic}For each graph and a sample of fractions of removed
arcs we show the change in harmonic diameter (by the measure $\delta$
defined in Section~\ref{sec:measures}) and the percentage of reachable pairs. PR stands
for PageRank, and LP for label propagation.}
\end{sidewaystable}

It is interesting to compare our results against those in the previous
literature. With respect to~\cite{AJBEATCN}, we test much larger networks. We
can confirm that random removal is less effective that rank-based removal, but
clearly the variation in diameter measured in~\cite{AJBEATCN} has been made
on a \emph{symmetrised} version of the web graph. Symmetrisation destroys much
of the structure of the network, and it is difficult to justify (you cannot
navigate links backwards). We have evaluated our experiment using the variation
in diameter instead of the variation in average distance (not shown here), but the
results are definitely inconclusive. The behaviour is wildly different even
between graphs of the same type, and shows no clear trend. This was expected, as
the diameter is defined by a maximisation property, so it is very unstable.

We also evaluated the variation in harmonic diameter (see
Table~\ref{tab:harmonic}), to compare our results with those of~\cite{FogWSBW}.
The harmonic diameter is very interesting, as it combines reachability and
distance. The data confirm what we already stated: web graphs react to removal
of 30\% of their arcs by label propagation by increasing their harmonic diameter
by an order of magnitude---something that does not happen with social networks.
Table~\ref{tab:harmonic} is even more striking than Table~\ref{tab:results} in
showing that label propagation selects highly disruptive nodes in web graphs.

Our criterion for node elimination is a threshold on the number of \emph{arcs}
removed, rather than nodes, so it is not possible to compare our results
with~\cite{FogWSBW} directly. However, for \texttt{.uk} PageRank at $\theta=0.01$
removes 648 nodes, which produced in the \texttt{.ie} graph a relative increment of 100\%, whereas
we find 14\%. This is to be expected, due to the very small size of the dataset
used in~\cite{FogWSBW}: experience shows that connectedness phenomena in web
graphs are very different in the ``below ten million nodes'' region.
Nonetheless, the growth trend is visibile in both cases. However, the experiments in \cite{FogWSBW} fail 
to detect both the disruptive behaviour at $\theta=.3$ and the striking difference in behaviour between
largest-degree and PageRank strategy.

\section{Conclusions and future work}

We have explored experimentally the alterations of the distance distribution of
some social networks and web graphs under different node-removal strategies.
We have confirmed some of the experimental results that appeared in the
literature, but at the same time shown some basic limitations of previous
approaches. In particular, we have shown for the first time that there is a
clear-cut structural difference between social networks and web
graphs\footnote{In this paper, like in all the other experimental research on
the same topic, conclusions about social networks should be taken with a grain
of salt, due to the heterogeneity of such networks and the lack of a large
repertoire of examples.}, and that it is important to test node-removal
strategies until a significant fraction of the arcs have been removed.

Probably the most important conclusion is that ``scale-free'' models,
which are currently proposed for both web graphs and social networks, do not to
capture this important difference: for this reason, they can only make sense as
long as they are adopted as baselines.


It might be argued that reachable pairs and distance
distributions are too coarse as a feature. Nonetheless, we believe that they are
the most immediate \emph{global} feature that are approachable computationally.
For instance, checking whether node removal alters the clustering coefficient
would not be so interesting, because the clustering coefficient of each node
depends only on the structure of the neighbourhood of each node. Thus, by
removing first the nodes with high coefficient it would be trivial to make the
clustering coefficient of the graph decrease quickly. Such trivial approaches
cannot possibly work with reachable pairs or with distance distributions because
they are properties that depend on the graph as a whole.

Finally, the efficacy of label propagation as a removal strategy suggests
that it may be very interesting to study it as a
form of \emph{ranking}: an open question is whether it could be useful, for
instance, as a query-independent ranking for information-retrieval applications.

\bibliography{biblio,socinfo,law}

\end{document}

%% file: avg.table
	\multirow{4}{*}{Amazon}
		& Random
	 & 0.002 & $(98\%)$ 
	 & 0.000 & $(93\%)$ 
	 & 0.023 & $(84\%)$ 
	 & 0.045 & $(84\%)$ 
	 & 0.056 & $(77\%)$ 
	 & 0.081 & $(66\%)$ 
\\
		& Degree
	 & 0.000 & $(105\%)$ 
	 & 0.010 & $(98\%)$ 
	 & 0.008 & $(90\%)$ 
	 & 0.018 & $(87\%)$ 
	 & 0.023 & $(86\%)$ 
	 & 0.070 & $(73\%)$ 
\\
		& PR
	 & -0.004 & $(95\%)$ 
	 & 0.008 & $(99\%)$ 
	 & 0.029 & $(99\%)$ 
	 & 0.034 & $(93\%)$ 
	 & 0.044 & $(91\%)$ 
	 & 0.092 & $(89\%)$ 
\\
		& LP
	 & 0.007 & $(104\%)$ 
	 & 0.025 & $(101\%)$ 
	 & 0.043 & $(84\%)$ 
	 & 0.081 & $(82\%)$ 
	 & 0.108 & $(79\%)$ 
	 & 0.149 & $(64\%)$ 
\\
\hline
	\multirow{4}{*}{Enron}
		& Random
	 & 0.007 & $(91\%)$ 
	 & 0.003 & $(84\%)$ 
	 & -0.007 & $(75\%)$ 
	 & -0.013 & $(65\%)$ 
	 & 0.003 & $(65\%)$ 
	 & 0.009 & $(57\%)$ 
\\
		& Degree
	 & -0.010 & $(81\%)$ 
	 & 0.009 & $(73\%)$ 
	 & 0.035 & $(65\%)$ 
	 & 0.047 & $(61\%)$ 
	 & 0.062 & $(53\%)$ 
	 & 0.119 & $(43\%)$ 
\\
		& PR
	 & 0.001 & $(92\%)$ 
	 & 0.041 & $(72\%)$ 
	 & 0.051 & $(55\%)$ 
	 & 0.066 & $(38\%)$ 
	 & 0.129 & $(34\%)$ 
	 & 0.184 & $(25\%)$ 
\\
		& LP
	 & -0.010 & $(90\%)$ 
	 & -0.028 & $(76\%)$ 
	 & -0.042 & $(70\%)$ 
	 & -0.061 & $(52\%)$ 
	 & -0.065 & $(52\%)$ 
	 & -0.057 & $(40\%)$ 
\\
\hline
	\multirow{4}{*}{Hollywood}
		& Random
	 & -0.005 & $(98\%)$ 
	 & 0.011 & $(101\%)$ 
	 & 0.005 & $(90\%)$ 
	 & 0.012 & $(86\%)$ 
	 & -0.003 & $(73\%)$ 
	 & 0.014 & $(72\%)$ 
\\
		& Degree
	 & 0.003 & $(101\%)$ 
	 & 0.008 & $(106\%)$ 
	 & 0.002 & $(103\%)$ 
	 & 0.012 & $(94\%)$ 
	 & 0.020 & $(105\%)$ 
	 & 0.021 & $(92\%)$ 
\\
		& PR
	 & 0.007 & $(104\%)$ 
	 & 0.006 & $(97\%)$ 
	 & 0.019 & $(101\%)$ 
	 & 0.026 & $(99\%)$ 
	 & 0.027 & $(95\%)$ 
	 & 0.046 & $(94\%)$ 
\\
		& LP
	 & -0.016 & $(91\%)$ 
	 & -0.036 & $(76\%)$ 
	 & -0.051 & $(63\%)$ 
	 & -0.062 & $(52\%)$ 
	 & -0.062 & $(48\%)$ 
	 & -0.058 & $(43\%)$ 
\\
\hline
	\multirow{4}{*}{LiveJournal}
		& Random
	 & -0.001 & $(97\%)$ 
	 & -0.001 & $(95\%)$ 
	 & 0.004 & $(86\%)$ 
	 & 0.012 & $(87\%)$ 
	 & 0.023 & $(77\%)$ 
	 & 0.027 & $(66\%)$ 
\\
		& Degree
	 & 0.009 & $(104\%)$ 
	 & 0.018 & $(97\%)$ 
	 & 0.026 & $(94\%)$ 
	 & 0.037 & $(103\%)$ 
	 & 0.051 & $(99\%)$ 
	 & 0.074 & $(90\%)$ 
\\
		& PR
	 & 0.002 & $(98\%)$ 
	 & 0.023 & $(101\%)$ 
	 & 0.041 & $(96\%)$ 
	 & 0.055 & $(93\%)$ 
	 & 0.075 & $(97\%)$ 
	 & 0.111 & $(90\%)$ 
\\
		& LP
	 & -0.004 & $(101\%)$ 
	 & -0.021 & $(83\%)$ 
	 & -0.022 & $(78\%)$ 
	 & -0.034 & $(69\%)$ 
	 & -0.026 & $(69\%)$ 
	 & -0.041 & $(56\%)$ 
\\
\hline
	\multirow{5}{*}{\texttt{.it}}
		& Random
	 & -0.012 & $(95\%)$ 
	 & 0.014 & $(87\%)$ 
	 & 0.007 & $(78\%)$ 
	 & 0.016 & $(69\%)$ 
	 & 0.017 & $(57\%)$ 
	 & 0.040 & $(45\%)$ 
\\
		& Degree
	 & -0.011 & $(97\%)$ 
	 & 0.031 & $(97\%)$ 
	 & 0.062 & $(87\%)$ 
	 & 0.068 & $(94\%)$ 
	 & 0.080 & $(80\%)$ 
	 & 0.127 & $(75\%)$ 
\\
		& PR
	 & 0.005 & $(101\%)$ 
	 & 0.096 & $(89\%)$ 
	 & 0.164 & $(69\%)$ 
	 & 0.244 & $(61\%)$ 
	 & 0.308 & $(52\%)$ 
	 & 0.447 & $(34\%)$ 
\\
		& LP
	 & 0.010 & $(85\%)$ 
	 & 0.213 & $(42\%)$ 
	 & 0.378 & $(22\%)$ 
	 & 0.487 & $(16\%)$ 
	 & 0.583 & $(11\%)$ 
	 & 0.793 & $(5\%)$ 
\\
		& Near-Root
	 & 0.025 & $(88\%)$ 
	 & 0.301 & $(58\%)$ 
	 & 0.330 & $(39\%)$ 
	 & 0.320 & $(36\%)$ 
	 & 0.344 & $(35\%)$ 
	 & 0.365 & $(28\%)$ 
\\
\hline
	\multirow{5}{*}{\texttt{.uk}}
		& Random
	 & 0.001 & $(99\%)$ 
	 & 0.003 & $(82\%)$ 
	 & 0.033 & $(83\%)$ 
	 & 0.037 & $(74\%)$ 
	 & 0.056 & $(69\%)$ 
	 & 0.062 & $(49\%)$ 
\\
		& Degree
	 & 0.005 & $(101\%)$ 
	 & 0.011 & $(104\%)$ 
	 & 0.003 & $(97\%)$ 
	 & -0.002 & $(94\%)$ 
	 & 0.013 & $(93\%)$ 
	 & 0.025 & $(98\%)$ 
\\
		& PR
	 & 0.023 & $(92\%)$ 
	 & 0.067 & $(80\%)$ 
	 & 0.100 & $(64\%)$ 
	 & 0.126 & $(55\%)$ 
	 & 0.175 & $(51\%)$ 
	 & 0.265 & $(38\%)$ 
\\
		& LP
	 & 0.051 & $(87\%)$ 
	 & 0.236 & $(39\%)$ 
	 & 0.272 & $(24\%)$ 
	 & 0.373 & $(18\%)$ 
	 & 0.439 & $(13\%)$ 
	 & 0.458 & $(6\%)$ 
\\
		& Near-Root
	 & 0.068 & $(80\%)$ 
	 & 0.244 & $(52\%)$ 
	 & 0.260 & $(48\%)$ 
	 & 0.261 & $(45\%)$ 
	 & 0.308 & $(45\%)$ 
	 & 0.278 & $(34\%)$ 
\\
\hline

%% file: harm.table
	\multirow{4}{*}{Amazon}
		& Random
	 & 0.036 & $(98\%)$ 
	 & 0.100 & $(93\%)$ 
	 & 0.215 & $(84\%)$ 
	 & 0.245 & $(84\%)$ 
	 & 0.397 & $(77\%)$ 
	 & 0.624 & $(66\%)$ 
\\
		& Degree
	 & -0.033 & $(105\%)$ 
	 & 0.033 & $(98\%)$ 
	 & 0.121 & $(90\%)$ 
	 & 0.197 & $(87\%)$ 
	 & 0.204 & $(86\%)$ 
	 & 0.473 & $(73\%)$ 
\\
		& PR
	 & 0.057 & $(95\%)$ 
	 & 0.035 & $(99\%)$ 
	 & 0.077 & $(99\%)$ 
	 & 0.125 & $(93\%)$ 
	 & 0.172 & $(91\%)$ 
	 & 0.234 & $(89\%)$ 
\\
		& LP
	 & -0.013 & $(104\%)$ 
	 & 0.028 & $(101\%)$ 
	 & 0.259 & $(84\%)$ 
	 & 0.321 & $(82\%)$ 
	 & 0.409 & $(79\%)$ 
	 & 0.795 & $(64\%)$ 
\\
\hline
	\multirow{4}{*}{Enron}
		& Random
	 & 0.132 & $(91\%)$ 
	 & 0.247 & $(84\%)$ 
	 & 0.397 & $(75\%)$ 
	 & 0.552 & $(65\%)$ 
	 & 0.572 & $(65\%)$ 
	 & 0.879 & $(57\%)$ 
\\
		& Degree
	 & 0.249 & $(81\%)$ 
	 & 0.435 & $(73\%)$ 
	 & 0.623 & $(65\%)$ 
	 & 0.800 & $(61\%)$ 
	 & 1.041 & $(53\%)$ 
	 & 1.703 & $(43\%)$ 
\\
		& PR
	 & 0.139 & $(92\%)$ 
	 & 0.472 & $(72\%)$ 
	 & 0.951 & $(55\%)$ 
	 & 1.758 & $(38\%)$ 
	 & 2.285 & $(34\%)$ 
	 & 3.741 & $(25\%)$ 
\\
		& LP
	 & 0.145 & $(90\%)$ 
	 & 0.311 & $(76\%)$ 
	 & 0.425 & $(70\%)$ 
	 & 0.833 & $(52\%)$ 
	 & 0.835 & $(52\%)$ 
	 & 1.388 & $(40\%)$ 
\\
\hline
	\multirow{4}{*}{Hollywood}
		& Random
	 & 0.032 & $(98\%)$ 
	 & 0.029 & $(101\%)$ 
	 & 0.125 & $(90\%)$ 
	 & 0.178 & $(86\%)$ 
	 & 0.373 & $(73\%)$ 
	 & 0.432 & $(72\%)$ 
\\
		& Degree
	 & 0.013 & $(101\%)$ 
	 & -0.042 & $(106\%)$ 
	 & -0.011 & $(103\%)$ 
	 & 0.087 & $(94\%)$ 
	 & -0.014 & $(105\%)$ 
	 & 0.128 & $(92\%)$ 
\\
		& PR
	 & -0.024 & $(104\%)$ 
	 & 0.055 & $(97\%)$ 
	 & 0.028 & $(101\%)$ 
	 & 0.049 & $(99\%)$ 
	 & 0.100 & $(95\%)$ 
	 & 0.138 & $(94\%)$ 
\\
		& LP
	 & 0.104 & $(91\%)$ 
	 & 0.281 & $(76\%)$ 
	 & 0.537 & $(63\%)$ 
	 & 0.814 & $(52\%)$ 
	 & 0.978 & $(48\%)$ 
	 & 1.256 & $(43\%)$ 
\\
\hline
	\multirow{4}{*}{LiveJournal}
		& Random
	 & 0.046 & $(97\%)$ 
	 & 0.059 & $(95\%)$ 
	 & 0.185 & $(86\%)$ 
	 & 0.162 & $(87\%)$ 
	 & 0.331 & $(77\%)$ 
	 & 0.587 & $(66\%)$ 
\\
		& Degree
	 & -0.026 & $(104\%)$ 
	 & 0.057 & $(97\%)$ 
	 & 0.114 & $(94\%)$ 
	 & 0.018 & $(103\%)$ 
	 & 0.075 & $(99\%)$ 
	 & 0.203 & $(90\%)$ 
\\
		& PR
	 & 0.041 & $(98\%)$ 
	 & 0.027 & $(101\%)$ 
	 & 0.090 & $(96\%)$ 
	 & 0.162 & $(93\%)$ 
	 & 0.129 & $(97\%)$ 
	 & 0.261 & $(90\%)$ 
\\
		& LP
	 & 0.007 & $(101\%)$ 
	 & 0.200 & $(83\%)$ 
	 & 0.287 & $(78\%)$ 
	 & 0.413 & $(69\%)$ 
	 & 0.443 & $(69\%)$ 
	 & 0.745 & $(56\%)$ 
\\
\hline
	\multirow{5}{*}{\texttt{.it}}
		& Random
	 & 0.069 & $(95\%)$ 
	 & 0.173 & $(87\%)$ 
	 & 0.313 & $(78\%)$ 
	 & 0.503 & $(69\%)$ 
	 & 0.803 & $(57\%)$ 
	 & 1.345 & $(45\%)$ 
\\
		& Degree
	 & 0.033 & $(97\%)$ 
	 & 0.077 & $(97\%)$ 
	 & 0.220 & $(87\%)$ 
	 & 0.192 & $(94\%)$ 
	 & 0.389 & $(80\%)$ 
	 & 0.584 & $(75\%)$ 
\\
		& PR
	 & -0.002 & $(101\%)$ 
	 & 0.268 & $(89\%)$ 
	 & 0.690 & $(69\%)$ 
	 & 1.057 & $(61\%)$ 
	 & 1.524 & $(52\%)$ 
	 & 3.221 & $(34\%)$ 
\\
		& LP
	 & 0.200 & $(85\%)$ 
	 & 1.885 & $(42\%)$ 
	 & 5.022 & $(22\%)$ 
	 & 8.178 & $(16\%)$ 
	 & 13.285 & $(11\%)$ 
	 & 34.809 & $(5\%)$ 
\\
		& Near-Root
	 & 0.186 & $(88\%)$ 
	 & 1.233 & $(58\%)$ 
	 & 2.415 & $(39\%)$ 
	 & 2.698 & $(36\%)$ 
	 & 2.867 & $(35\%)$ 
	 & 3.763 & $(28\%)$ 
\\
\hline
	\multirow{5}{*}{\texttt{.uk}}
		& Random
	 & 0.024 & $(99\%)$ 
	 & 0.232 & $(82\%)$ 
	 & 0.245 & $(83\%)$ 
	 & 0.416 & $(74\%)$ 
	 & 0.538 & $(69\%)$ 
	 & 1.158 & $(49\%)$ 
\\
		& Degree
	 & -0.005 & $(101\%)$ 
	 & -0.022 & $(104\%)$ 
	 & 0.046 & $(97\%)$ 
	 & 0.072 & $(94\%)$ 
	 & 0.092 & $(93\%)$ 
	 & 0.046 & $(98\%)$ 
\\
		& PR
	 & 0.122 & $(92\%)$ 
	 & 0.340 & $(80\%)$ 
	 & 0.716 & $(64\%)$ 
	 & 1.053 & $(55\%)$ 
	 & 1.320 & $(51\%)$ 
	 & 2.238 & $(38\%)$ 
\\
		& LP
	 & 0.216 & $(87\%)$ 
	 & 2.117 & $(39\%)$ 
	 & 4.152 & $(24\%)$ 
	 & 6.231 & $(18\%)$ 
	 & 9.229 & $(13\%)$ 
	 & 22.050 & $(6\%)$ 
\\
		& Near-Root
	 & 0.339 & $(80\%)$ 
	 & 1.327 & $(52\%)$ 
	 & 1.572 & $(48\%)$ 
	 & 1.712 & $(45\%)$ 
	 & 1.844 & $(45\%)$ 
	 & 2.663 & $(34\%)$ 
\\
\hline